# Topological Dirac-vortex modes in a three-dimensional photonic topological insulator

Bei Yan[1,2,#], Yingfeng Qi[1,#], Ziyao Wang[1,#], Yan Meng[3], Linyun Yang[4], Zhen-Xiao Zhu[1], Jing-Ming Chen[1], Yuxin Zhong[1], Min-Qi Cheng[1], Xiang Xi[3,*], Zhen Gao[1,§]

[1]State Key Laboratory of Optical Fiber and Cable Manufacturing Technology, Department of Electronic and Electrical Engineering, Guangdong Key Laboratory of Integrated Optoelectronics Intellisense, Southern University of Science and Technology, Shenzhen 518055, China
[2]Hubei Province Key Laboratory of Systems Science in Metallurgical Process, and College of Science, Wuhan University of Science and Technology, Wuhan 430081, China
[3]School of Electrical Engineering and Intelligentization, Dongguan University of Technology, Dongguan, 523808, China
[4]College of Aerospace Engineering, Chongqing University, Chongqing, 400030, China
[#]These authors contributed equally to this work.
[*,§]Corresponding author. Email: xix@dgut.edu.cn (X.X.); gaoz@sustech.edu.cn (Z.G.)

**Recently, topological Dirac-vortex modes in Kekulé-distorted photonic lattices have attracted broad interest and exhibited promising applications in robust photonic devices such as topological cavities, lasers, and fibers. However, due to the vectorial nature of electromagnetic waves that results in complicated band dispersions and fails the tight-binding model predictions, it is challenging to construct three-dimensional (3D) topological photonic structures with Kekulé distortion and the photonic topological Dirac-vortex modes have thus far been limited to two-dimensional (2D) systems. Here, by directly mapping a 3D Kekulé-distorted tight-binding model in a 3D tight-binding-like photonic crystal exhibiting scalar-wave-like band structures, we theoretically propose and experimentally demonstrate topological Dirac-vortex modes in a 3D photonic topological insulator for the first time. Using microwave near-field measurements, we directly observe robust photonic topological Dirac-vortex modes bound to and propagate along a one-dimensional (1D) Dirac-vortex line defect, matching well with the tight-binding and simulation results. Our work offers an ideal platform to map tight-binding models in 3D topological photonic crystals directly and opens a new avenue for exploiting topological lattice defects to manipulate light in 3D space.**

**Introduction**

Recently, the interplay between real-space topological lattice defects (TLD) [1, 2] and reciprocal-space band topology [3] has given rise to many novel topological phenomena and promising applications [4-6], such as cavities [7-15], lasers [16-20], waveguides [21-30], fibers [31-32], and three-dimensional (3D) photonic topological insulators in synthetic dimensions [33]. In particular, topological Dirac-vortex modes bound to a vortex defect in Kekulé-distorted lattices, which are Jackiw-Rossi zero modes originating from the Dirac equation with mass vortices [34], have attracted widespread attention in many areas ranging from high-energy physics [35], condensed-matter physics [36] to topological physics [12-20, 32, 37-42] due to its unique properties such as scalable mode areas, arbitrary mode degeneracies, vector-beam emission, and large free spectral range. However, to date, most previous studies of topological Dirac vortex modes have been limited to two-dimensional (2D) systems that support zero-dimensional (0D) localized modes.

More recently, one-dimensional (1D) vortex-string chiral modes bound to and propagate along a Dirac-vortex line defect [41] and 0D monopole topological modes localized in a 3D Dirac-vortex volume defect [42] have been experimentally demonstrated in 3D acoustic crystals, extending topological Dirac-vortex modes from 2D to 3D systems. However, in sharp contrast to the studies of topological Dirac-vortex modes in 3D acoustic crystals [41-42] where the couplings and on-site energies can be engineered flexibly to implement the discrete lattice models (tight-binding models) easily and directly, their photonic counterparts have been severely lagged due to the vectorial nature of electromagnetic waves and the inherent challenge of discretely modeling the 3D photonic systems [43-44], which usually makes the tight-binding model prediction fail. Therefore, it is still an open question whether topological Dirac-vortex modes can be realized in 3D topological photonic structures (even in theory).

Here, we theoretically propose and experimentally demonstrate topological Dirac-vortex modes in a 3D photonic topological insulator [45-49] by inducing Kekulé distortion in a 3D tight-binding-like photonic crystal [50-52] whose bulk band dispersions resemble scalar waves and match well with that of the 3D tight-binding model. Using microwave near-field measurements, we directly observe topological Dirac-vortex modes bound to and propagate along the 1D Dirac-vortex line defect in a 3D photonic topological insulator. Moreover, we experimentally demonstrate that the photonic topological Dirac-vortex modes exhibit superior robustness against various obstacles, making such modes well-suited for robust electromagnetic wave manipulation in 3D space. Our work not only experimentally extends photonic topological Dirac-vortex modes from 2D to 3D for the first time, but also provides a versatile platform to explore novel physical phenomena and practical applications enabled by TLDs in 3D topological photonic crystals.

**Results**

**Topological Dirac-vortex modes in a three-dimensional Kekulé-distorted honeycomb lattice**

We start with topological Dirac-vortex modes in a 2D Kekulé-distorted honeycomb lattice, as schematically illustrated in the upper panel of Fig. 1a, where the topological Dirac-vortex modes (red region) are tightly localized around the 0D vortex core due to the gapped Dirac cones induced by mass terms with phase vortices. The winding number of the Dirac vortices equals the number of topological Dirac-vortex modes. Then we vertically stack the 2D Kekulé-distorted honeycomb lattices with uniform interlay couplings to construct a 3D Kekulé-distorted honeycomb lattice which supports 1D topological Dirac-vortex modes (red arrows) bound to and propagate along the 1D vortex line defect, as shown in the lower panel of Fig. 1a. To study this unique 1D topological Dirac-vortex modes, we adopt a 3D tight-binding model whose unit cell consists of two layers of honeycomb lattices with intralayer couplings $t_a$ (green rods), $t_b$ (blue rods), $t_c$ (orange rods), and interlayer couplings $t_z$ (grey rods), respectively, as shown in Fig. 1b (see the Hamiltonian in Supplementary Information). The intralayer coupling strengths are $t_a = t_0 - \delta t \cos(\varphi + 4\pi/3)$, $t_b = t_0 - \delta t \cos(\varphi + 2\pi/3)$, and $t_c = t_0 - \delta t \cos(\varphi)$, where $t_0$ represents the initial coupling strength, $\delta t$ and $\varphi$ represent the amplitude and phase of the Dirac mass, respectively. When all intralayer couplings are equal ($t_a = t_b = t_c$, $\delta t = 0$), the bulk band structure (green lines) of the 3D honeycomb lattice in the first Brillouin zone (BZ) (Fig. 1c) is gapless with an eightfold degenerate double Dirac point at A point, as shown in Fig. 1d. When we introduce periodic Kekulé distortion characterized by different intralayer couplings ($t_a \neq t_b \neq t_c$, $\delta t \neq 0$ and $\varphi = \pi/3$), the eightfold degenerate double Dirac point will be broken, resulting in a gapped bulk band structure (grey lines) with a complete 3D topological bandgap (orange region) and topological surface/hinge states (see the Wilson loop and eigenstate calculation in Supplementary Information). To study the nontrivial topological states supported by a 3D aperiodic Kekulé-distorted honeycomb lattices characterized by a position-dependent phase modulation $\varphi(r)$ to the 3D honeycomb lattice, we calculate the dispersion of a finite hexagonal supercell of 3D aperiodic Kekulé-distorted honeycomb lattice with open boundary conditions in the *xy* plane and periodic boundary conditions along the *z* direction, as shown in Fig. 1e, the 3D aperiodic Kekulé-distorted honeycomb lattice supports topological Dirac-vortex modes (red line) and topological hinge states (blue line) within the 3D topological bandgap (orange region). The topological hinge states and the topological Dirac-vortex modes can be selectively excited by placing a point source (green star) at the hinge or center of the 3D Kekulé-distorted honeycomb lattice, respectively, as shown in Figs. 1f-1g, we can see that the topological hinge states and topological Dirac-vortex modes are tightly localized and propagate along the hinge or vortex line defect, respectively. Since the topological hinge states have been experimentally demonstrated in a 3D photonic higher-order topological insulator [53], we only focus on the 1D topological Dirac-vortex modes in this work.

**Topological Dirac-vortex modes in a 3D photonic topological insulator**

Now we design a 3D photonic topological insulator with Kekulé distortion by implementing the 3D tight-binding model (lower panel of Fig. 1a) in a 3D tight-binding-like photonic crystal with confined Mie resonance [50]. Figs. 2a-2b show the perspective and top views of a unit cell of the 3D tight-binding-like photonic crystal, in which the twelve dielectric rods (white rods) serve as sites. Each dielectric rod is embedded by three metallic rods (copper rods) to confine the Mie resonance to satisfy the tight-binding approximation in photonic systems. The perforated metallic plates with air holes are used to introduce interlayer couplings. Remarkably, the embedded metallic rods and perforated metallic plates confine the Mie resonances of the dielectric rods, making the complex vectorial electromagnetic waves in 3D photonic crystals simplified to scalar-wave-like ones and creating chiral symmetric photonic band structures that ideally match those of the tight-binding models with nearest-neighbor couplings. The Kekulé distortion can be introduced in the 3D tight-binding-like photonic crystal by displacing the dielectric rods from their original positions with distance $m_0$ and angle $\varphi$, as indicated in Fig. 2b. Similar to the tight-binding model, when $m_0 = 0$ and $\varphi = \pi/3$, the simulated bulk band structure of the 3D tight-binding-like photonic crystal (green lines) exhibits a double Dirac point with eightfold degeneracy at the A point, and the degeneracy will be lifted (grey lines) and open a 3D complete topological photonic bandgap (orange region) when $m_0 \neq 0$ and $\varphi = \pi/3$, as shown in Fig. 2c. More significantly, if $m_0 \neq 0$, the 3D photonic bandgap persists for arbitrary angle of $\varphi$, as shown in Fig. 2d, in which the color represents the angle $\varphi$ varying from 0 to $2\pi$ (see the band inversion in Supplementary Information).

We then implement aperiodic Kekulé distortion to a 3D tight-binding-like photonic crystals to open a varying vortex bandgap by defining a modulation vector (Dirac mass) $\boldsymbol{m} = m_0 e^{jw\varphi(\boldsymbol{r})}$ with a winding number $w = 1$ and position-dependence phase $\varphi(\boldsymbol{r}) = \arg(\boldsymbol{r})$ varying continuously from 0 to $2\pi$, as schematically shown in Fig. 2e. Fig. 2f presents the simulated dispersion of the topological Dirac-vortex mode (red lines) along the $k_z$ direction within the minimum vortex bandgap ranging from 18.5 to 19 GHz (orange region). Fig. 2g shows the $E_z$ field distribution of the topological Dirac-vortex eigenmode marked by a black dot in Fig. 2f, we can see that the topological Dirac-vortex mode is tightly localized around the vortex core. It is worth noting that the number of topological Dirac-vortex modes is determined by the Dirac-mass winding number $w$ and multiple topological Dirac-vortex modes, having nearly identical group and phase velocities, can exist simultaneously with large Dirac-mass winding numbers (see Supplementary Information for details). We employ a point dipole source (green star) to excite the topological Dirac-vortex modes in a 3D Kekulé-distorted photonic topological insulator, as shown in Fig. 2h, in which we can see that the topological Dirac-vortex modes are bound to and propagate bidirectionally along the 1D vortex line defect.

**Experimental observation of topological Dirac-vortex modes in a 3D photonic topological insulator**

Next, we experimentally demonstrate the topological Dirac-vortex modes in a 3D tight-binding-like photonic crystal. The fabricated experimental sample is shown in Fig. 3a, which consists of forty layers (20 unit cells) of perforated copper plates with air holes and perforated air foams inserted with dielectric and metallic rods. The large and small air holes in the perforated copper plates are used to introduce interlayer couplings and insert probe antenna to map the topological Dirac-vortex modes, respectively. Figs. 3b-3c shows the top view of a perforated air foam inserted with dielectric and metallic rods. We first measure the transmission spectra of the topological Dirac-vortex modes (red line) and bulk states (grey line) by placing a point source antenna (green star) at the center of the sample and inserting a probe antenna into the top center (red star) or top boundary (grey star) of the sample, as shown in Fig. 3d, in which we can see that the transmission of the topological Dirac-vortex modes (red line) within the bulk bandgap (orange region) is much higher than that of the bulk states (grey line). To directly observe the topological Dirac-vortex modes, we insert a probe antenna into the small air holes one by one to map the $E_z$ field distribution of the topological Dirac-vortex modes at 18.7 GHz, as shown in Fig. 3e, the topological Dirac-vortex modes are bound to and propagate vertically along the 1D vortex line defect, agreeing well with the simulation results (Fig. 2h) and unambiguously verifying the existence of topological Dirac-vortex modes in the 3D Kekulé-distorted photonic crystals. The unique propagation characteristic of the topological Dirac-vortex modes can also be revealed by the measured electric field distributions as a function of the excitation frequencies and the $z$ coordinates of the probe antenna, as shown in Fig. 3f. For the frequency range of 18.5-19 GHz within the bandgap, the topological Dirac-vortex modes can propagate upward and downward simultaneously, indicating broadband propagation of topological Dirac-vortex modes along the vortex line defect in the bandgap. Meanwhile, the vanished upward and downward transmissions outside the bandgap indicate the absence of topological Dirac-vortex modes along the vortex line defect. Moreover, by Fourier-transforming the complex electric field distributions from real space to reciprocal space, we can obtain the measured dispersion (color map) of the topological Dirac-vortex modes, as shown in Fig. 3g, which matches well with the simulation results (cyan solid lines).

Finally, we examine the robustness of the topological Dirac-vortex modes. We introduce local defects (green dashed square) in the upper center of the vortex line defect by either removing the central six dielectric rods or replacing them with six metallic rods (blue circles), as shown in Fig. 4a and Fig. 4b, respectively. We then measure the transmission spectra of the topological Dirac vortex modes with (blue line) and without (red line) local defects, as shown in Fig. 4c and Fig. 4d, respectively, we can see that they almost overlap with each other, indicating the robustness of the topological Dirac-vortex modes despite the presence of various defects. We repeat the near-field imaging measurements to directly observe the propagation of the topological Dirac-vortex modes in the presence of local defects, as shown in Fig. 4e and Fig. 4f, respectively, the topological Dirac-vortex modes can circumvent the defects (green dashed square) and continue to propagate along the 1D vortex line defect. For comparison, we present the simulated electric distributions of the topological Dirac-vortex modes with defects, as shown in Fig. 4g and Fig. 4h, which agree well with the experimental results.

**Discussion**

In conclusion, by directly emulating a 3D Kekulé-distorted tight-binding model in a 3D tight-binding-like photonic crystal, we have theoretically proposed and experimentally demonstrated 1D topological Dirac-vortex modes in a 3D photonic topological insulator for the first time. We also experimentally observed that the photonic topological Dirac-vortex modes are robust against defects or obstacles, making them suitable for robust manipulation of electromagnetic waves in 3D space. Moreover, we show that the 3D tight-binding-like photonic crystal exhibits scalar-wave-like band dispersions resembling those of the tight-binding model, making the experimental realization of 3D photonic topological phases an easy task. We envision other topological defects such as dislocation, disclination, and monopole topological modes that can be readily realized in the 3D tight-binding-like photonic crystals.

**Methods**

**Numerical simulations.** All numerical results presented in this work are simulated using the RF module of COMSOL Multiphysics. The bulk band structures are calculated using a unit cell with periodic boundary conditions in all directions. The perforated copper plates and metallic rods are modeled as perfect electric conductors (PEC) in the simulation. The dispersion of the topological Dirac-vortex modes is calculated by adopting a hexagonal supercell and applying periodic boundary conditions along the $z$ direction, and open boundary conditions along the $x$ and $y$ directions. In the full-wave simulations of a finite 3D tight-binding-like photonic crystal, all boundaries are set as open boundary conditions.

**Materials and experimental setups.** The copper plates are fabricated by depositing a 0.035 mm-thick layer of copper onto a Teflon woven-glass fabric laminate. We use perforated air foam (ROHACELL 31 HF with a relative permittivity of 1.04 and a loss tangent of 0.0025) to fix the metallic and dielectric rods. In the experimental measurements, the amplitude and phase of the electric fields are measured using a vector network analyzer (Keysight E5080) connected by two electric dipole antennas serving as the source and probe, respectively. To excite the topological Dirac-vortex modes, a point source antenna is placed at the center of the sample, and a probe antenna is inserted into the air holes one by one to scan the electric fields.

**Data availability**

The data that support the findings of this study are available from the corresponding authors upon reasonable request.

**Code availability**

We use commercial software COMSOL Multiphysics to perform electromagnetic numerical simulations. Requests for computation details can be addressed to the corresponding authors.

**Acknowledgments**

Z.G. acknowledges the finding from the National Natural Science Foundation of China under grants No. 62361166627, 62375118, and 12104211, Guangdong Basic and Applied Basic Research Foundation under grant No.2024A1515012770, Shenzhen Science and Technology Innovation Commission under grants No. 20220815111105001 and 202308073000209, High level of special funds under grant No. G03034K004. Y. M acknowledges the support from the National Natural Science Foundation of China under Grant No. 12304484, and the Guangdong Basic and Applied Basic Research Foundation under grant No. 2024A1515011371

**Authors Contributions**

Z.G. initiated and supervised the project. B.Y., Y.F.Q., Z.Y.W., and X.X. performed the simulations. B.Y., Y.F.Q, Z.Y.W., X.X., and Z.G. designed the experiments. B.Y., Z.G., Z.Y.W., Y.F.Q., X.X., L.Y., Y.M., Z.X.Z., Y.X.Z., M.Q.C., and J.M.C. fabricated samples. B.Y., Y.F.Q., and Z.Y.W. carried out the measurements. B.Y., Y.F.Q., Z.Y.W., X.X., and Z.G. analyzed data. B.Y. and Z.G. wrote the manuscript.

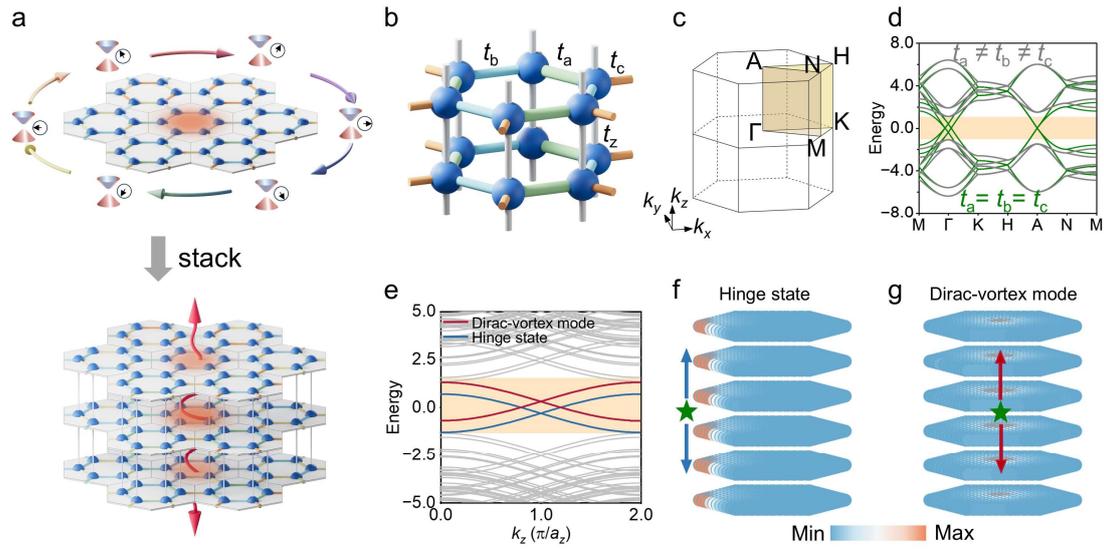

**Fig. 1 | Topological Dirac-vortex modes in 3D Kekulé-distorted honeycomb lattice. a** Upper panel: 2D Kekulé-distorted honeycomb lattice hosting 0D localized topological Dirac-vortex modes (red region). Lower panel: 3D Kekulé-distorted honeycomb lattice supporting 1D propagating topological Dirac-vortex modes (red arrow). **b** Unit cell of the 3D honeycomb lattice with intralayer ($t_a$, $t_b$, and $t_c$) and interlayer ($t_z$) couplings. **c** 3D Brillouin zone. **d** Calculated bulk band structures of the 3D honeycomb lattice without (green lines) and with (grey lines) Kekulé modulation, respectively. The orange region represents the 3D topological energy bandgap. **e** Calculated dispersions of topological Dirac-vortex modes (red lines) and hinge states (blue lines), and the grey lines and orange region represent the bulk states and 3D topological energy bandgap, respectively. **f, g** Energy distributions of the topological hinge states (**f**) and Dirac-vortex modes (**g**), respectively. The green arrow represents a point source to excite the topological hinge states and Dirac-vortex modes.

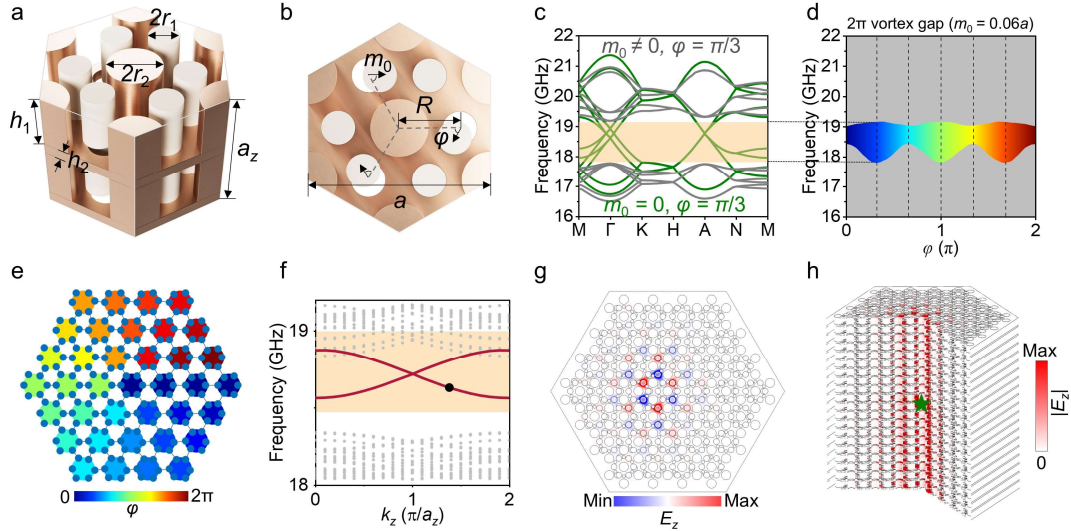

**Fig. 2 | Topological Dirac-vortex modes in a 3D photonic topological insulator**. **a** Schematic of the unit cell of a 3D tight-binding-like photonic crystal. The copper parts represent the perforated metallic plates and metallic rods, and the white parts represent the dielectric rods. **b** Top view of the unit cell. The lattice constants in the $xy$ plane and $z$ direction are $a = 15$ mm and $a_z = 11.7$ mm, respectively. The other geometrical parameters are $R = 5$ mm, $r_1 = 1.5$ mm, $r_2 = 2.4$ mm, $m_0 = 0.9$ mm, $h_1 = 1$ mm, $h_2 = 4.85$ mm, respectively. The Kekulé distortion is introduced by displacing the dielectric rods with distance $m_0$ and angle $\varphi$. **c** Simulated bulk band structures without (green lines) and with (grey lines) Kekulé distortion, respectively. The orange region represents the 3D topological photonic bandgaps. **d** 3D topological photonic bandgaps with different angles $\varphi$ and fixed displacement distance $m_0 = 0.06a$. The color represents the angle $\varphi$ varying from 0 to $2\pi$. **e** Schematic of the modulation angle $\varphi$ of the aperiodic Kekulé distortion with a winding number of +1. **f** Simulated dispersion of the topological Dirac-vortex modes (red lines) along the $k_z$ direction, the orange region represents the minimum Dirac-vortex bandgap. **g** Simulated electric field distribution of the topological Dirac-vortex eigenmodes at $k_z = 1.5\pi/a_z$ marked by the black dot in **f**. **h** Simulated electric field distribution of the topological Dirac-vortex modes excited by a point source (green star) at 18.7 GHz.

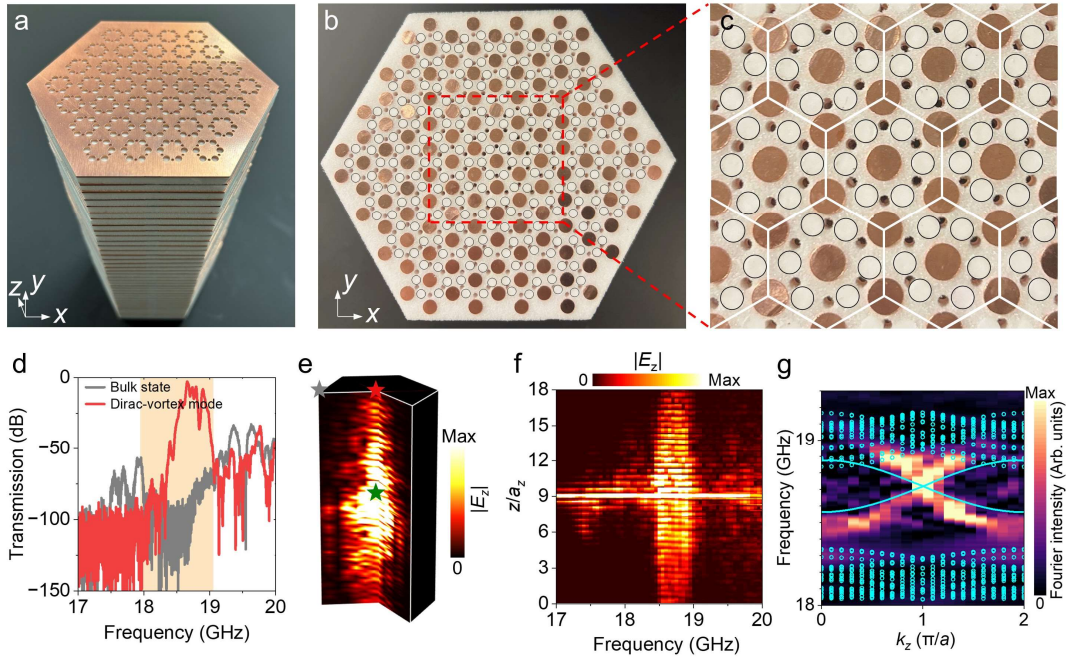

**Fig. 3 | Experimental observation of topological Dirac-vortex modes in a 3D photonic topological insulator. a** Photograph of the fabricated experimental sample consisting of 40 layers of perforated copper plates and air foams inserted with dielectric and metallic rods. **b** Photograph of a perforated air foam inserted with metallic (copper part) and dielectric (black circles) rods. **c** Magnified image of the perforated air foam. **d** Measured transmission spectra of the bulk (grey line) and topological Dirac-vortex (red line) modes, the orange region represents the 3D photonic bandgap. **e** Measured electric field distribution of the topological Dirac-vortex mode excited by a point source (green star) at 18.7 GHz. The green (red and grey) star represents the point source (probe antennas). **f** Measured electric field distributions as a function of the excitation frequencies and the $z$ coordinates of the probe antenna. **g** Measured (color map) and simulated (cyan solid lines) dispersions of the topological Dirac-vortex modes.

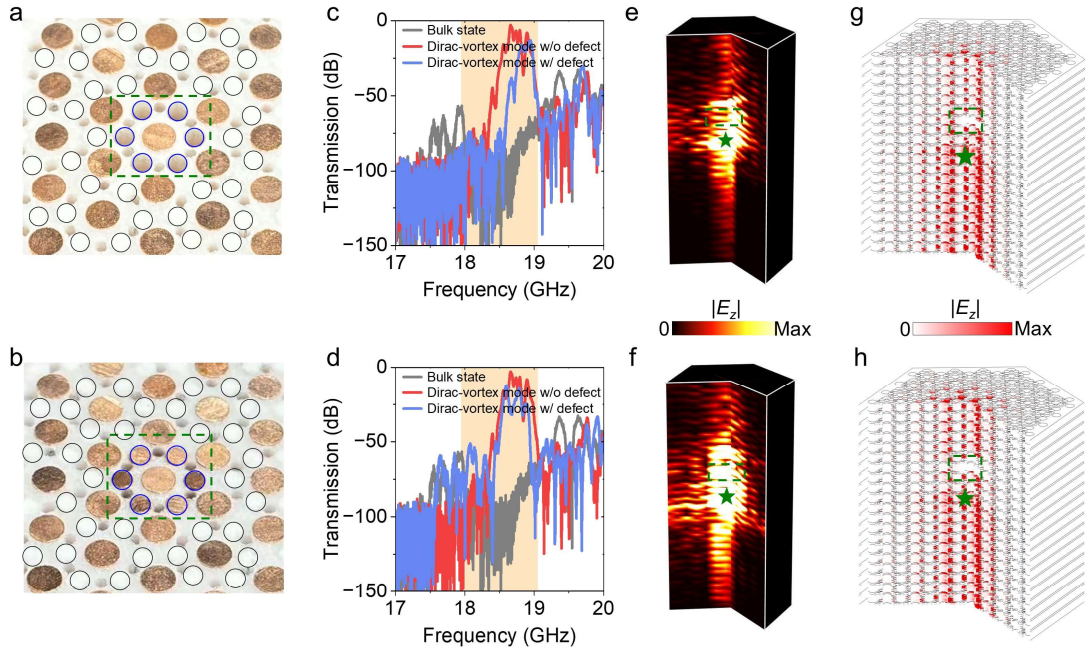

**Fig. 4 | Robustness of the photonic topological Dirac-vortex modes. a, b** Photographs of the local defects by removing six dielectric rods (**a**) or replacing them with six metallic rods (**b**) at the upper center of the vortex line defect. **c**, **d** Measured transmission spectra of the topological Dirac vortex modes with (blue lines) and without (red lines) local defects in **a** and **b**, respectively. **e**, **f** Measured electric field distributions of the topological Dirac vortex modes with six dielectric rods removed (**e**) or replaced by six metallic rods (**f**) at 18.7 GHz, respectively. **g**, **h** Simulated electric field distributions of the topological Dirac vortex modes with six dielectric rods removed (**e**) or replaced by six metallic rods (**f**) at 18.7 GHz, respectively. The green star and green dashed square represent the point source and local defects, respectively.